# Universality of 2d Yukawa and Gross–Neveu models

Erich Focht[a] *

[a]Institute for Theoretical Physics E, RWTH-Aachen, 52074 Aachen, Germany
HLRZ c/o KFA Jülich, 52425 Jülich, Germany

Evidence for the same universal behavior of 2d Yukawa ($Y_2$) and Gross–Neveu (GN) models in a certain range of couplings, particularly also for $\kappa < 0$, is presented.

## 1. Introduction

Yukawa models in 2d ($Y_2$) with chiral Z(2) and U(1) symmetry are investigated on the lattice. They are natural extensions of the GN models, which are specially interesting because of their asymptotic freedom and the dynamical mass generation. Some years ago the equivalence between $Y_2$ and GN models has been suggested in [1] using a mean field (MF) analysis. Recently we have found numerical evidence for the asymptotic freedom of the $Y_2$ models in a large range of couplings [2,3]. This property is suggested in the effective potential approach [4] and by our MF method [3], which is reliable in 2d due to infrared divergence of the fermion loop inducing long-range effective scalar field interaction.

In this paper we discuss the question whether $Y_2$ and GN models are in the same universality class for a certain range of couplings. Most of the results were obtained by simulating the action with staggered fermions and hypercubic Yukawa coupling, e.g. in the Z(2) case:

$$S = \sum_x \left[ -2\kappa \sum_\mu \phi_x \phi_{x+\mu} + \phi_x^2 + \lambda(\phi_x^2 - 1)^2 \right]$$
$$+ \frac{1}{2} \sum_{x,\alpha,\mu} \left( \overline{\chi}_x^\alpha \eta_{\mu x} \chi_{x+\mu}^\alpha - \overline{\chi}_{x+\mu}^\alpha \eta_{\mu x} \chi_x^\alpha \right)$$
$$+ y \sum_{x,\alpha} \overline{\chi}_x^\alpha \phi_x^P \chi_x^\alpha \qquad (1)$$

where $\kappa$ is the hopping parameter, $\lambda$ the quartic selfcoupling of the scalar field, $\eta_{\mu x}$ the sign factor

*Work done in collaboration with W.Franzki, J.Jersák, M.Stephanov. Supported by D.F.G. and B.M.F.T.

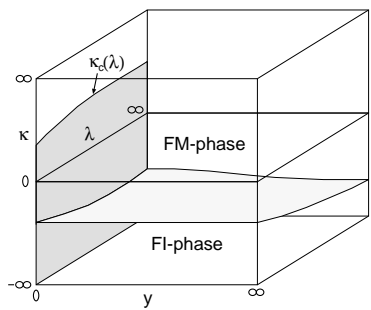

Figure 1. Sketch of the phase diagram of $Y_2$ models with *hypercubic* Yukawa coupling.

for staggered fermions and $\phi_x^P$ the average of $\phi$ on the hypercube with the lower left corner $x$. The number of fermions in the continuum is $N_F = 4N$, $\alpha = 1..N$ labeling the fermion species.

In this representation the GN model arises in the limit $\kappa = \lambda = 0$, while at $\lambda = \infty$, $y = 0$ eq. (1) describes the Ising or the XY model, $\kappa_c(\lambda)$ being the critical line.

## 2. Properties of $Y_2$ models

Works on the GN models and on the $Y_2$ models [1–4] lead us to the phase diagramm sketched in fig. 1. It is based on $1/N$ expansions, which can be justified only for $\lambda = O(1/N)$, MF approximations, which can give very good results in 2d [5], and numerical simulations of our group.

Due to the logarithmic divergence in 2d, the phase diagramm differs from that of higher dimensions. At small $y$ there is no paramagnetic phase in the Z(2) case, the whole sheet $\{y = 0, \kappa \leq \kappa_c(\lambda)\}$ being critical. At $\kappa \approx -\kappa_c(\lambda)$ another critical surface separates the ferromagne-tic (FM) from the ferrimagnetic (FI) phase. In the U(1) case the phase structure is similar, we have above $-\kappa_c(\lambda)$ the spin wave (SW), below the staggered spin wave phase.

Fermion mass is generated dynamically in both $Y_2$ models [3,5,6] for arbitrarily small $y$. The continuum limit can be performed by approaching the critical surface at $y = 0$, $\kappa < \kappa_c(\lambda)$.

$Y_2$ models are asymptotically free like the GN models [7,2,3]. At fixed $\kappa$ and $\lambda$, for $\kappa < \kappa_c(\lambda)$ the MF approximation [3,5] suggests in the continuum limit the scaling law:

$$am_F \sim \exp\left(-\frac{1}{2\beta_0}\frac{1}{\chi(\kappa,\lambda)y^2}\right), \qquad (2)$$

$\chi(\kappa,\lambda)$ being the susceptibility in the scalar theories ($y = 0$) and $\beta_0$ the first coefficient of the $\beta$-function in the GN models. For $\kappa > 0$ it is $\chi = Z_\phi/a^2 m_\phi^2$, but the MF prediction holds also for $\kappa < 0$.

## 3. Universality

The phase structure in fig. 1 suggests a classification of the possible universality classes in the $Y_2$ models, depending on where and how the continuum limit is performed. At $y = 0$ we have the usual $\phi^4$ theories, the phase transition at $\kappa_c(\lambda)$ is for $\lambda > 0$ Ising like in the Z(2), Kosterlitz–Thouless (XY) like in the U(1) case. The Gaussian fixed point is at $\lambda = 0$, $\kappa = 1/4$.

When approaching the critical line $\kappa_c(\lambda)$ from positive $y$, both, fermions and bosons will remain in the spectrum and the situation is very interesting. The mass ratio of fermions and scalar particles must depend on the way the continuum limit is performed, characterizing a whole family of universality classes, similar to the case described by A. Sokal at this conference [8].

At the critical surface around $\kappa \simeq -\kappa_c(\lambda)$, $y > 0$, the fermion mass doesn't scale. In the corresponding continuum limit only the staggered scalars would remain in the spectrum.

In the continuum limit performed on the critical surface $y = 0$, $\kappa < \kappa_c(\lambda)$ only the fermions and their bound states survive in the spectrum. Fermions scale like in the GN model, but it is not clear whether GN and $Y_2$ models belong to the same universality class in this region. Two aspects of this question will be discussed in the following: whether $\beta_0$ of $Y_2$ and GN models is the same, and whether the mass ratio of the first two states is the same.

The first coefficient of the $\beta$ function can be interpreted as a critical exponent. To compare it with that of the GN model one has to consider the GN scaling law: $am_F \sim \exp\left(-\frac{1}{2\beta_0}\frac{1}{g^2}\right)$, with $g$ being the usual 4 fermion coupling. The $\beta_0$ coefficients would be the same if the effective scaling variable in the $Y_2$ case is $g = \sqrt{\chi}y$. This is indeed suggested by perturbation theory in $y$. The fermion 4-point function is modified already on the tree level in $y$ by the scalar field. At zero momentum it is:

$$\Gamma_F^{(4)}(p_i = 0) = y^2 G_S^{(2)}(p_i = 0) = y^2 \chi(\kappa,\lambda). \qquad (3)$$

As eq. (2) is valid only in infinite volume at very small $y$, we have to modify it in order to compare it with finite lattice simulations [3]. We use lattice sums instead of momentum integrals and the full response function $f(H)$ determined in numerical simulations in order to make predictions about $am_F$ with the MF method. The selfconsistency eq.

$$\langle\phi\rangle = f[Nyam_F \sum_{\{p\}}(\sum_\mu \sin^2 p_\mu + a^2 m_F^2)^{-1}], \quad (4)$$

with $am_F = y\langle\phi\rangle$, must be solved numerically, by recursion. The result is compared to $am_F$. But as this method is only an approximation comparable to 1 fermion loop results, we have to consider eventual corrections to it. Therefore we introduce a parameter which quantifies the deviation from the MF prediction by taking $am_F = r \cdot y\langle\phi\rangle$ and fitting several $am_F$ values with the function obtained by solving (4). Fig. 2 shows a typical fit and the resulting values of $r$ in such fits are compiled in **table 1**.

It is clear that $r \approx 1$ means that the scaling of $am_F$ is GN like. But certain deviations can be expected, because MF is not perfect. An estimate for these deviations is given by the fit in the GN case, where $r$ differs from 1 by about 5%. Such a deviation for other $\kappa$, $\lambda$ values thus does not contradict the hypothesis, that $\beta_0$ is the same as in the GN case.

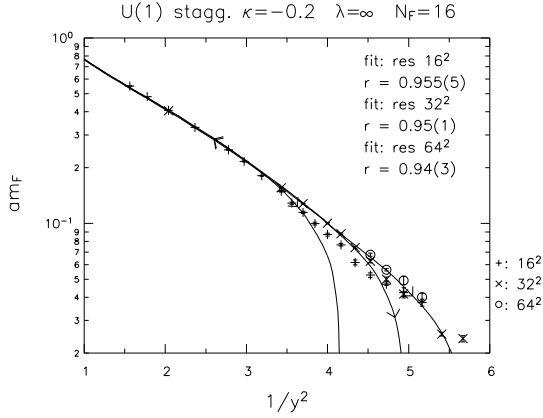

Figure 2. Scaling behavior of $am_F$ for $\kappa = -0.2$ and $\lambda = \infty$ on $16^2$, $32^2$ and $64^2$ lattices. The lines are fits for the parameter $r$ defined in the text.

We decided to calculate the $r$ values for $\lambda = \infty$, at maximal bare scalar selfcoupling. The results for $\kappa = -0.4$ and $\kappa = -0.2$ show that $r \approx 1$ whithin the limits allowed by the GN value at $\kappa = \lambda = 0$. This strongly supports the assumption that $Y_2$ and GN models are equivalent **even at negative $\kappa$**.

| $\kappa$ | $\lambda$ | $16^2$ | $32^2$ | $64^2$ |
|---|---|---|---|---|
| 0 | 0 | 0.957(3) | 0.956(5) | |
| -0.4 | $\infty$ | 0.971(5) | 0.970(7) | |
| -0.2 | $\infty$ | 0.955(5) | 0.95(1) | 0.94(3) |
| 0 | $\infty$ | 0.94(1) | 0.93(2) | 0.93(4) |
| 0.2 | $\infty$ | 0.91(4) | 0.86(4) | 0.85(6) |

Table 1: The values of the parameter $r$.

At $\kappa = 0.2$ the parameter $r$ differs considerably from 1. The quality of the fits was poor, i.e. the data could not be described well by the MF approach. This is probably an effect of the crossover to the universality classes at $\kappa = \kappa_c(\lambda)$. It could also explain the relatively large deviation ($r \approx 0.93$) at $\kappa = 0$, $\lambda = \infty$.

The mass ratios are universal quantities and should be the same if both models are in the same universality class. We tried to measure the first one. In fig. 3 $m_{\overline{F}F}/m_F$ is plotted against another dimensionless quantity, $m_F L$. Data from all lattices with $L/a = 8, 16, 32, 64$ is plotted with the same symbols, so the mass ratios seem to scatter more than they in fact do on a given lattice.

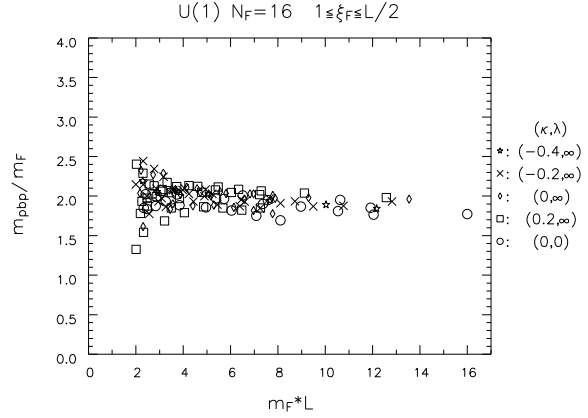

Figure 3. The mass ratio $m_{\overline{F}F}/m_F$ vs. $m_F L$.

All $\kappa$ values give the same results and they are consistent with those of the GN model (circles).

Concluding we can say that we found substantial evidence for the assumption that $Y_2$ and GN models are in the same universality class for $\kappa < \kappa_c(\lambda)$. Particularly interesting is the fact that this holds also for negative $\kappa$, which is beyond the scope of the considerations in continuum [1].

In 2d Yukawa theories reduce to the simpler GN models. The situation is different from that in 4d, where the Yukawa theories are those with the right number of parameters, not the 4 fermion (Nambu – Jona-Lasinio) models [9].